\documentclass[twocolumn,floats,floatfix,prd,aps,superscriptaddress,showpacs,showkeys,nobalancelastpage]{revtex4}
\usepackage{amsmath,amsfonts,amsthm,amssymb}
\usepackage[pdftex]{graphicx}
\usepackage{hyperref}

\newcommand{\unit}[1]{\ensuremath{\, \mathrm{#1}}}

\newcommand{\inner}[2]{\ensuremath{\left( #1 \left| #2 \right)\right.}}
\newcommand{\expt}[1]{\ensuremath{\left\langle #1\right\rangle}}
\newcommand{\infint}{\int_{-\infty}^{\infty}}
\newcommand{\halfint}{\int_{0}^{\infty}}
\newcommand{\Dt}{\Delta t}
\newcommand{\Deff}{\ensuremath{D_{\rm eff}}}

\begin{document}
\title{Summed Parallel Infinite Impulse Response (SPIIR) Filters For
  Low-Latency Gravitational Wave Detection}
\author{Shaun Hooper}
\email{shaun.hooper@uwa.edu.au}
\affiliation{Australian International Gravitational Research Centre,  School of Physics, University of Western Australia, 35 Stirling Hwy, Crawley, WA 6009, Australia}
\affiliation{ICRAR-Fairway M468, School of Physics, The University of Western Australia, Crawley, WA 6009, Australia}
\author{Shin Kee Chung}
\affiliation{Australian International Gravitational Research Centre,  School of Physics, University of Western Australia, 35 Stirling Hwy, Crawley, WA 6009, Australia}
\affiliation{ICRAR-Fairway M468, School of Physics, The University of Western Australia, Crawley, WA 6009, Australia}
\author{Jing Luan}
\affiliation{Theoretical Astrophysics 350-17, California Institute of Technology, Pasadena, CA 91125, USA}
\author{David Blair}
\affiliation{Australian International Gravitational Research Centre,  School of Physics, University of Western Australia, 35 Stirling Hwy, Crawley, WA 6009, Australia}
\author{Yanbei Chen}
\affiliation{Theoretical Astrophysics 350-17, California Institute of Technology, Pasadena, CA 91125, USA}
\author{Linqing Wen}
\email{linqing.wen@uwa.edu.au}
\affiliation{Australian International Gravitational Research Centre,  School of Physics, University of Western Australia, 35 Stirling Hwy, Crawley, WA 6009, Australia}
\affiliation{ICRAR-Fairway M468, School of Physics, The University of Western Australia, Crawley, WA 6009, Australia}


\begin{abstract}
  With the upgrade of current gravitational wave detectors, the first
  detection of gravitational wave signals is expected to occur in the next
  decade. Low-latency gravitational wave triggers will be necessary to make
  fast follow-up electromagnetic observations of events related to their
  source, e.g., prompt optical emission associated with short gamma-ray
  bursts. In this paper we present a new time-domain low-latency algorithm for
  identifying the presence of gravitational waves produced by compact binary
  coalescence events in noisy detector data. Our method calculates the signal
  to noise ratio from the summation of a bank of parallel infinite impulse
  response (IIR) filters. We show that our \emph{summed parallel infinite
    impulse response} (SPIIR) method can retrieve the signal to noise ratio to
  greater than 99\% of that produced from the optimal matched filter. We
  emphasise the benefits of the SPIIR method for advanced detectors, which
  will require larger template banks.
\end{abstract}

\maketitle
\section{Introduction}

The interferometric gravitational wave (GW) detectors LIGO \cite{LIGOweb}, and
Virgo \cite{VIRGOweb} have reached a sensitivity at which the detection of GWs
is possible. The LIGO detectors are currently undergoing a major upgrade to
Advanced LIGO, for which the sensitivity will be improved ten fold relative to
Initial LIGO \cite{Smith2009}. Hence Advanced LIGO will be able to detect GW
(GW) sources within a volume of space one thousand times larger than that of
initial LIGO, out to $\sim$200-300\unit{Mpc} \cite{AdvancedLIGOdesign}.

The emission of GWs produced by compact binary coalescence (CBC) can be
modelled with a high degree accuracy \cite{Abbott2009b}. When two compact
bodies, such as neutron stars or black holes are in orbit, Einstein's
equations predict the generation of GWs. As the bodies spiral towards each
other a GW is created that increases in frequency over time until the bodies
merge, following what is known as the inspiral waveform. Ground based
detectors have frequency passbands that allow them to be sensitive to the
final stages of such events up to a total system masses of several hundred
$M_{\odot}$.

Neutron star binary mergers are widely thought to be the progenitors of short
hard gamma-ray bursts (short GRBs) \cite{Fox2005, Nakar2007}. The delay between
the final GW emission and the onset of the GRB is estimated to be as short as
0.1 seconds or as long as tens to hundreds of seconds \cite{vanPutten2009,
  Zhang2004}. The electromagnetic emission of the GRB event is not well
understood. Related to the initial GRB there is thought to be a prompt
emission in the X-ray and optical wavelengths followed by a delayed afterglow
of cascading wavelengths. Prompt optical emission may occur tens to hundreds
of seconds after the initial burst. The low-latency detection of the GW
associated with a neutron star merger could lead to the localisation of a GRB
source event on the sky, enabling fast moving telescopes to observe the prompt
optical emission. Data collected from a multitude of sources --- GWs,
gamma-rays, X-rays and optical counterparts of the GRB --- will lead to maximum
insight into these highly energetic events.

The standard strategy for searching for the existence of inspiral waveforms in
the detector data is based on matched filtering \cite{Abbott2009b} (and
references therein). This method, based on Wiener optimal filtering, is a
correlation of an expected inspiral waveform template and the detector data,
weighted by the inverse noise spectral density of the detector
\cite{WandZ}. In order to save computational costs, this correlation is
performed in the frequency domain, via a Fourier transform of a finite segment
of detector data. In previous LIGO searches, the detector data is split up
into ``science blocks'', which are further divided into ``data segments''
chosen to be at least twice the length of the longest waveform in the template
bank \cite{Abbott2008}. Each proceeding data segment is chosen to overlap the
previous one by 50\%. Each segment therefore must be matched filtered in a
time that is half the length of the segment for a real-time analysis, that
  is, the filter output rate is equal to the data input rate. In this case,
the matched filter process has a minimum latency (from signal arrival to
signal detection) that is proportional to the longest template (see
\cite{Luan2011} for more details). Advanced LIGO will have an increased
bandwidth over Initial LIGO, with the lower bound dropping from 40\unit{Hz} to
10\unit{Hz} \cite{AdvancedLIGOdesign}. GW signals from CBC events spend much
more time at these lower frequencies. Hence waveforms used for matched
filtering in Advanced LIGO will be much longer (1000s of seconds). This in
turn means the segment length will be increased, further increasing the
latency. The latency of this method to produce GW triggers is longer than the
time to onset of prompt optical emission after coalescence (10s to 100s of
seconds). After this amount of time, the early electromagnetic counterpart of
a GRB event will be significantly faded, and may be missed by telescopes
altogether.

A low-latency GW detection method is required to trigger follow-up
electromagnetic observations of the prompt optical emission. So far two
frequency domain methods have been developed to solve this issue. The VIRGO
group has produced a low-latency pipeline based on \emph{Multi-Band Template
  Analysis} (MBTA) \cite{Buskulic2010}, and LIGO is also working on a new
method, \emph{Low-Latency On-line Inspiral Data} analysis (LLOID) method. In
MBTA the matched filtering technique is split over two frequency bands, and
the output is coherently added, reducing latency. A latency of less than 3
minutes until the availability of a trigger using this method has been
achieved \cite{Buskulic2010}. Low-latency in the LLOID method is achieved by
first down-sampling the incoming data into multiple streams and then applying
frequency domain finite impulse response (FIR) filters \cite{gstlal}. The
computational cost of this pipeline is reduced by decreasing the number of
templates via singular value decomposition \cite{Cannon2010}.

\begin{figure}[htb]
  \centering
  \includegraphics[width=2in]{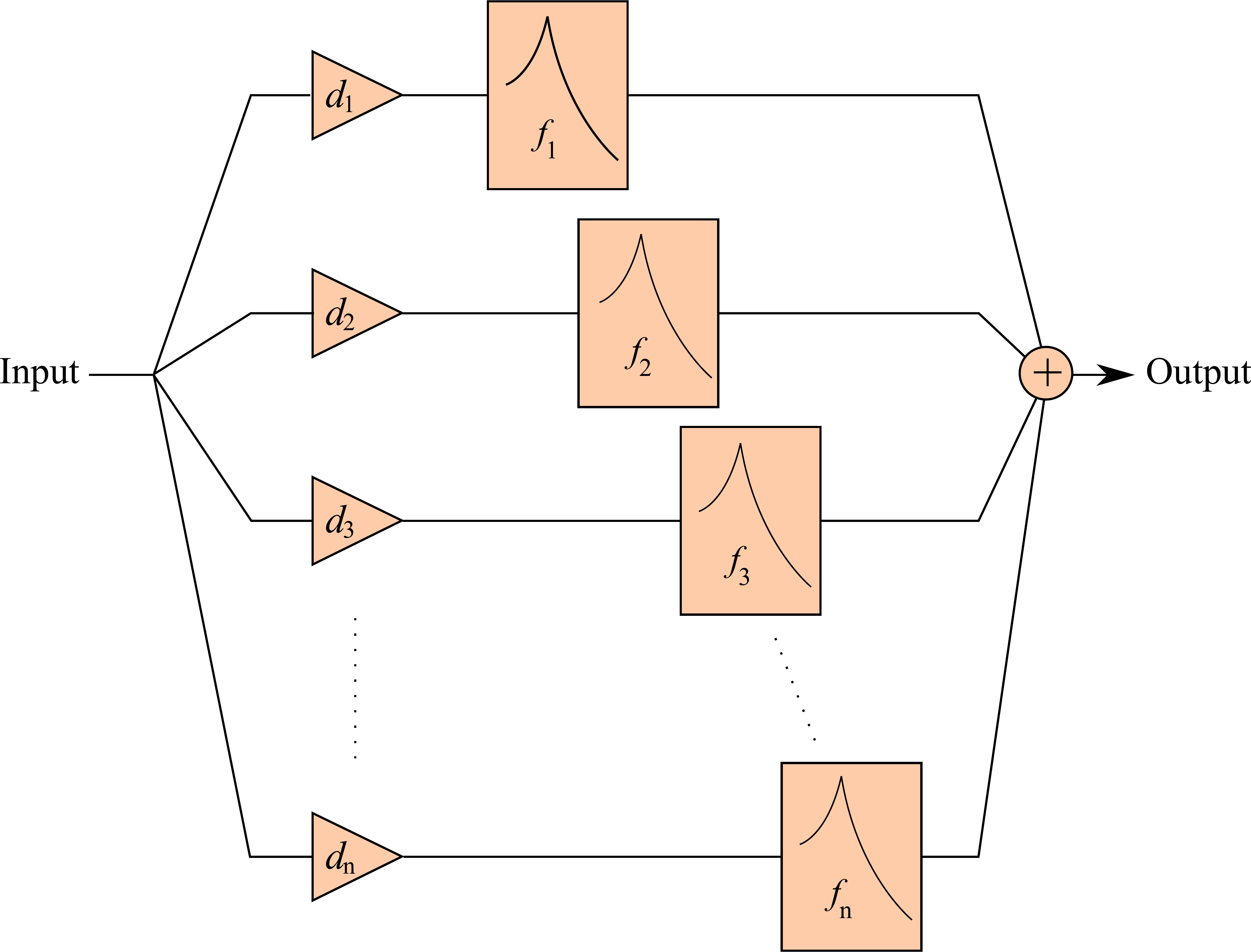}
  \caption{A schematic overview of the SPIIR method. The input is split into
    different channels, time delayed by an amount $d$, then passed through a
    narrow bandpass IIR filters, each with a different central frequency
    $f$. Finally the output of each individual IIR filter is summed, giving
    the output of the SPIIR method.}
  \label{fig:bandpassbank}
\end{figure}

We introduce a new method to detect CBC signals in the time domain using
infinite impulse response (IIR) filters. Approximating an inspiral waveform by
a summation of time shifted exponentially increasing sinusoids enables us to
construct a bank of parallel single-pole IIR filters. Each IIR filter acts as
a narrow bandpass filter. When each appropriately delayed IIR filter is added
the coherent output approximates the matched filter output of the exact
waveforms. We call this the \emph{summed parallel infinite impulse response}
(SPIIR) method. Figure \ref{fig:bandpassbank} visually demonstrates the idea
of using a bank of IIR filters as narrow bandpass filters.
For a full explanation of the mathematical principles, see \cite{Luan2011}. In
this follow up paper, we numerically address the issues essential to the
practical use of this method for the upcoming advanced detectors. We calculate
the filter coefficients and demonstrate via numerical simulations how well our
method approximates the optimal matched filter as a function the number of
filters per bank using a range of parameters. We also show that the detection
rate of the SPIIR method is very similar to that of the matched filter
method. It has been shown theoretically that in order to get the same latency
as the SPIIR method, the frequency domain matched filter method would require
greater computational resources \cite{Luan2011}.

The structure of this paper is as follows: In section \ref{sec:details} we
will go through the formal introduction of the inspiral waveform and matched
filtering, and how to get from the continuous frequency domain matched filter
to the time domain discrete matched filter. This will lead to a demonstration
on how it is possible to approximate an inspiral signal by a sum of
exponentially increasing sinusoids. The methodology is explained in Section
\ref{sec:method} and will cover how we set up our simulation to test the
efficiency of the SPIIR method as opposed to the frequency domain matched
filter. Section \ref{sec:results} will analyse the results of the simulation
and Section \ref{sec:summary} will discuss the implications of these results
for advanced detectors.

\section{Methodology}\label{sec:details}

Gravitational wave interferometers output the strain induced by gravitational
waves incident on the detector, as well as inherent noise. In unitless
strain, the detector output will be,
\begin{align}
  s(t) = \begin{cases}
    n(t) & \text{if signal is absent} \\
    n(t) + h(t) & \text{if signal is present} \end{cases}
\end{align}
where $n(t)$ is the noise inherent in the detector. The sensitivity of the
instrument can be characterized by the (one-sided) strain power spectral
density $S_n(f)$,
\begin{align}
  \expt{\tilde{n}(f)\tilde{n}^*(f')} = \frac{1}{2}S_n(f)\delta(f-f')
\end{align}
where the tilde represents the forward Fourier transform,
\begin{align}
  \tilde{q}(f) = \infint q(t) e^{-2\pi i f t} dt.
\end{align}

\subsection{The Inspiral Waveform}

The gravitational-wave strain incident at the interferometer is given by
\begin{align}
  h(t) = F_+(\theta,\phi,\psi)h_+(t) + F_{\times}(\theta,\phi,\psi)h_{\times}(t) \label{eq:strain}
\end{align}
where the detector response functions $F_+$ and $F_{\times}$ are functions of
$(\theta,\phi)$ - the standard spherical polar coordinates measured with
respect to the Earth's fixed frame, and $\psi$ is the polarisation angle. The
detector response function can be found in \cite{Anderson2001}. The $+$ and
$\times$ polarisations of the waveform are,
\begin{align}
  h_+(t) &= \left(\frac{1+\cos^2\iota}{2}\right)A(t)\cos\phi(t) \label{eq:hplus}\\
  h_{\times}(t) &= \left(\cos\iota\right) A(t)\sin\phi(t)\label{eq:hcross}
\end{align}
For non-spinning binaries with a chirp mass \mbox{$\mathcal{M} =
  ((m_1m_2)^3/(m_1+m_2))^{1/5}$} in the range of $1-3M_{\odot}$ --- we will
hereafter assume --- the waveforms can be modelled to very high accuracy using
the Restricted post-Newtonian (PN) expansion \cite{Abbott2004,Blanchet1995a,
  Blanchet1996} in the LIGO band (assumed to be 10-1500 \unit{Hz} for advanced
LIGO). For restricted waveforms, only the leading order of the amplitude
$A(t)$ is taken,
\begin{align}
  A(t) &= \frac{G\mathcal{M}}{Dc^2}\left(\frac{t_c-t}{5G\mathcal{M}/c^3}\right)^{-1/4}
\end{align}
and the post-Newtonian phase $\phi(t)$ is given by
\begin{align}
  \phi(t) &= \phi_c-2\left(\frac{t_c - t}{G\mathcal{M}/c^3}\right)^{5/8} + \mbox{higher order terms}
\end{align}

In addition to the source masses $m_1,m_2$, there are several unknown
parameters; the time of coalescence $t_c$, the phase at coalescence $\phi_0$,
distance from observer to source $D$, the inclination angle of the binary's
orbital plane relative the line of sight $\iota$, and the polarisation angle
$\psi$. However by using the linear combination trigonometric identity, one can
re-express the strain \eqref{eq:strain} by splitting the scaling factor due to
distance, sky location and orientation to the mass dependant time evolution of
the waveform \cite{Fairhurst2008},
\begin{align}
  h(t) &= \frac{1\unit{Mpc}}{\Deff} \left[ h_c(t)\cos\phi_0 + h_s(t)\sin\phi_0\right] \label{eq:waveform}
\end{align}
where the scalar factor $\Deff$ is,
\begin{align}
  \Deff = \frac{D}{\sqrt{F_+^2\left(1+\cos^2 \iota\right)^2/4
      +F_{\times}^2\left(\cos\iota\right)^2}}
\end{align}
which gives $\phi_0$, an unknown phase as,
\begin{align}
  \phi_0 = \phi_c + \arctan\frac{F_{\times}\left(2\cos\iota\right)}{F_+\left(1+\cos^2\iota\right)}
\end{align}
We now define the terms $h_c$ and $h_s$ as the waveform at $\phi_0=0$ and
$\phi_{\pi/2}$, scaled at 1\unit{Mpc} as the so called ``cosine'' and ``sine'' phases
\cite{Brady2008},
\begin{align}
  h_c(t) &= A_{1\unit{Mpc}}(t)\cos\phi(t) \label{eq:cospol}\\
  h_s(t) &= A_{1\unit{Mpc}}(t)\sin\phi(t) \label{eq:sinpol}
\end{align}

\subsection{The Matched Filter}\label{sec:matchedfilter}

The matched filter $Q$ is a linear operator that maximises the ratio of
``signal'' to ``noise'' present in the detector data $s$ \cite{Allen2005}. It
is denoted by,
\begin{align}
  z(t) &= 2 \infint \frac{\tilde{s}(f)\tilde{Q}^*(f)}{S_n(|f|)}e^{2\pi i ft}
  df = \inner{s(t)}{Q} \label{eq:matchedfilter}
\end{align}
Where we have also defined the inner product \inner{a}{b}. The signal to noise
ratio (SNR) is \emph{generally} defined as the ratio of observed filter output
to it's expected root-mean square flucations or standard deviation,
\begin{align}
  \mathrm{SNR} &= \frac{z}{\sqrt{\expt{(z-\expt{z})^2}}} = \frac{z}{\sqrt{\expt{z}^2}} = \frac{z}{\sqrt{\inner{Q}{Q}}}
\end{align}
Note that in the absence of a signal, $\expt{\mathrm{SNR}}=0$ and
$\expt{(\mathrm{SNR})^2}=1$ \textbf{independent} of the normalisation of the
filter $Q$.

\subsection{Two-Phase Filter}

A convenient way to search for the unknown phase constant $\phi_0$ is to filter
both phases $h_c$ and $h_s$ separately and then combined to form a
complex signal. The two-phase filter is defined as,
\begin{subequations}\label{eq:twophasef}
\begin{align}
  z(t)  &= \inner{s(t)}{h_c} + i\inner{s(t)}{h_s}\\
  \begin{split}&=2\infint \frac{\tilde{s}(f)\tilde{h}_c^*(f)}{S_n(f)}e^{2\pi i ft} df \\
    & \qquad \qquad+ i2\infint\frac{\tilde{s}(f)\tilde{h}_s^*(f)}{S_n(|f|)}e^{2\pi i ft} df \end{split}
\end{align}
\end{subequations}
The advantage of using the phases $h_{c,s}$ is that in the stationary
phase approximation \cite{Droz1999}, $h_c$ and $h_s$ are exactly orthogonal
($\inner{h_c}{h_s}=\inner{h_s}{h_s}$, $\inner{h_c}{h_s}=0$). It then follows,
$\tilde{h}_{c}(f) = i\tilde{h}_{s}(f)$ for $f > 0 $. Generally, this is
applied to \eqref{eq:twophasef} to give the two-phase matched filter as,
\begin{align}
  z(t) &= 4 \halfint \frac{\tilde{s}(f)\tilde{h}_{c}^*(f)}{S_n(|f|)}e^{2\pi i ft} df \label{eq:zcomplex}
\end{align}
However in this paper, we prefer to maintain the form of the two-phase filter
in \eqref{eq:twophasef}. In convention with the field, the amplitude signal to
noise ratio of the (quadrature) matched filter is defined as the absolute
value of the two-phase filter, divided by a normalisation constant that is
equal to standard deviation of the real and imaginary parts of the two-phase
filter,
\begin{align}
  \rho(t) = \frac{|z(t)|}{\sigma} \label{eq:rho}
\end{align}
where $\sigma^2$ is,
\begin{align}
  &\sigma^2 = 2\infint \frac{\left|\tilde{h}_c(f)\right|^2}{S_n(f)} df = \inner{h_c}{h_c}
\end{align}
Note that in the in the absence of a signal (just noise), the SNR $\rho$
\eqref{eq:rho} is Rayleigh distributed with mean $\sqrt{\pi/2}$ and variance
$1$, which is identical to the Chi-distribution with two degrees of freedom
(one for each of the phases). This of course implies that the SNR squared,
$\rho^2$ is Chi-square distributed with two degrees of freedom. Hence the
probability of finding an SNR value greater than $\rho_*$ is \cite{Brady2008},
\begin{align}
  P(\rho^2>\rho_*^2) = e^{-\rho_*^2 / 2}.
\end{align}

\subsection{Digital Time Domain Filtering}

The two-phase matched filter \ref{eq:twophasef} is a cross correlation of
phase $h_{c,s}(t)$ and the detector output $s(t)$, weighted by the inverse
noise spectral density $S_n(f)$. By defining the quantity $x$ as the
\emph{over}-whitened strain data,
\begin{align} \label{eq:whiten}
  x(t) = \infint \frac{\tilde{s}(f)}{S(f)}e^{2\pi i f t} df
\end{align}
we can use the cross-correlation theorem to define the two-phase matched
filter in the time domain,
\begin{align}
  z(t) &= 2\int_{-\infty}^t x(t') h_c(t'-t) dt' + i2\int_{-\infty}^t x(t')
  h_s(t'-t) dt' \\
  &= 2\int_{-\infty}^tx(t') \hat{h}(t'-t) dt' \label{eq:zcorr}
\end{align}
where $\hat{h} = h_c(t) + i h_s(t) = A(t)e^{i\phi(t)}$.

The discrete form of the continuous time domain matched filter
\eqref{eq:zcorr} is,
\begin{equation}
  z_k = 2 \sum_{j=-\infty}^{k} x_j\hat{h}_{j-k} \Delta t \label{eq:zdiscrete}
\end{equation}
where $t = k \Dt$. In practise, the inspiral waveform template $h_i$ is
bounded (because the detector is only sensitive over a bandwidth), and the
summation becomes finite, making this a \emph{finite impulse response} (FIR)
filter.

\subsection{Infinite Impulse Response Filter}

Now let us introduce an alternative digital filter, the \emph{infinite impulse
  response} (IIR) filter. The difference equation of a general IIR filter is,
\begin{align}
  y_k = \sum_{n=1}^N a_n y_{k-n} + \sum_{m=0}^M b_m x_{k-m}
\end{align}
where $y_k$ is the filter output at time step $k$, ($t=k \Delta t$), $x_k$ is
the filter input, and $a$'s and $b$'s are complex coefficients.

Examples of IIR filters in common usage are Chebyshev, Butterworth and
elliptic filters. IIR filters use much less computational resources than an
equivalent FIR filter. This is because they have ``memory'' --- the previous
outputs are fed back into the filter. However \emph{digital} IIR filter design
is a more complex process than FIR design. Obtaining the coefficients is
usually done by first constructing an equivalent analog filter and applying
well-known methods, such as the bi-linear transform or impulse
invariance. Multiple IIR filters used together have different forms, such as
direct form I \& II, cascade (series) and parallel. In a series configuration,
the overall transfer function is the multiplication of each IIR filter
transfer function. In a parallel bank of IIR filters, where the output is
summed together, the overall transfer function is the summation of the
different transfer functions.

First, let's analyse the simplest single-pole IIR filter. The difference
equation of this filter is
\begin{equation}
  y_k = a_1y_{k-1} +b_0x_{k}. \label{eq:IIRdiff}
\end{equation}
\begin{figure}[htb]
  \centering
  \includegraphics[width=\columnwidth]{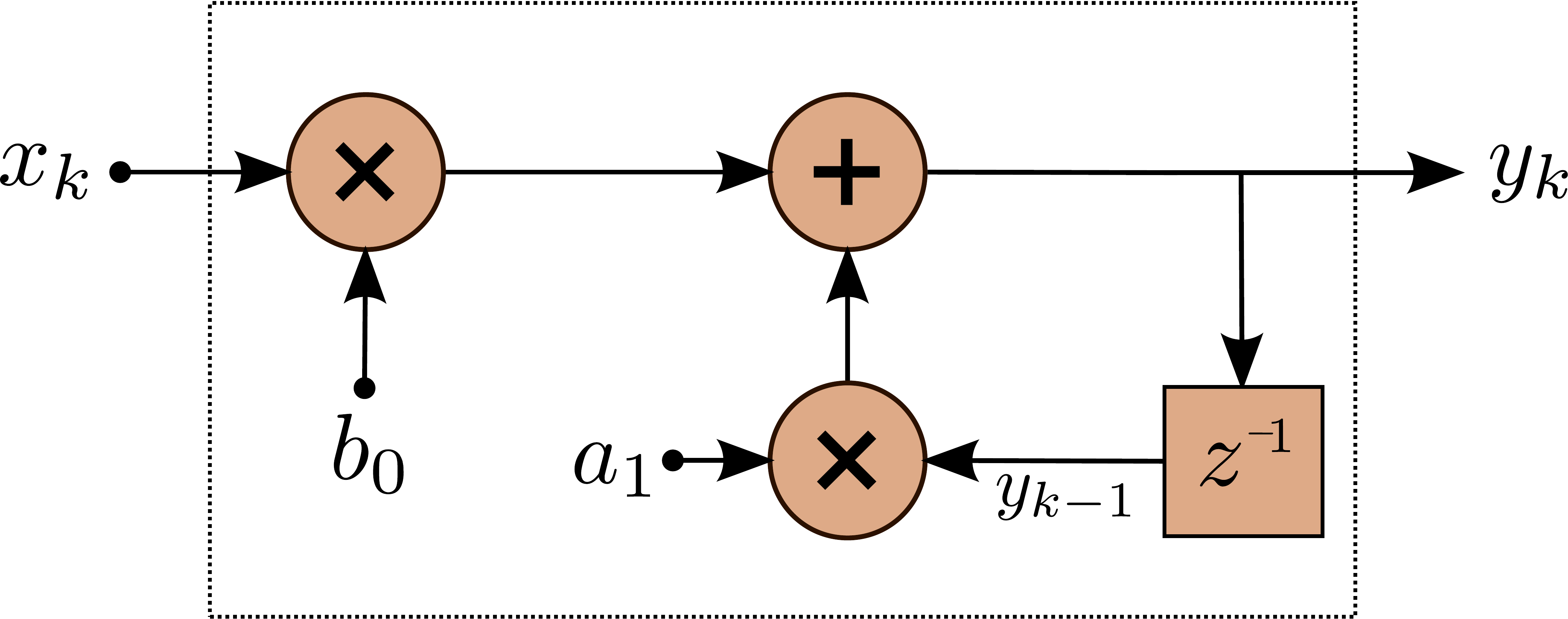}
  \caption{A signal processing schematic showing the flow of data through a
    digital single-pole IIR filter. The input, $x_k$ is multiplied by a
    complex constant $b_0$, then added to the previous output that has been
    multiplied by another complex constant $a_1$, resulting in the current
    output $y_k$. It should be noted that this filter, in principle, should be
    have been run forever.}
  \label{fig:iirfilter}
\end{figure}
A solution to this first-order linear inhomogeneous difference equation is
\begin{align}
  y_k = \sum_{j=-\infty}^{k} x_j b_0 a_1^{k-j}.  \label{eq:diffsoln}
\end{align}
By defining the complex coefficient $a_1$ in the form,
\begin{align}
  a_1 = e^{-(\gamma + i\omega)\Dt}
\end{align}
and comparing \eqref{eq:zdiscrete} and \eqref{eq:diffsoln}, it is easy to see
that the output of the simple filter \eqref{eq:IIRdiff} is the
cross-correlation of $x_k$ and complex sinusoid $u_n$ with frequency $\omega$
and a magnitude that increases with an exponential factor $\gamma$ for $n<0$:
\begin{align}
  u_n = b_0e^{(\gamma + i\omega)n\Dt}\Theta(-n) \label{eq:sinusoid}
\end{align}
where $\Theta(-n)$ is the Heaviside function.

\subsection{Approximation to an inspiral waveform} \label{sec:approx}

Since $\phi(t)$ is not linear in time, a complex sinusoid \eqref{eq:sinusoid}
cannot approximate the $h_{c,s}$ phases of the inspiral waveform
\mbox{$\hat{h}(t) = A(t)e^{i\phi(t)}$}. However we can easily linearise the
phases by a first-order Taylor expansion about the time $t_l^*$:
\begin{align}
  A(t)e^{i\phi(t)} \simeq A(t_l^*) e^{i\phi(t_l^*) + i\dot{\phi}(t_l^*)(t-t_l^*)};
\end{align}
since the amplitude $A(t)$ does not increase at the same rate as $\phi(t)$,
only a linear expansion of $\phi(t)$ is required. Multiplying by the window
function $e^{\gamma_l(t-t_l)}\Theta(t_l-t)$ makes this approximation an
exponentially increasing constant frequency complex sinusoid with cutoff time
$t_l$:
\begin{align}
  u_l(t)&=A(t_l^*)
  e^{i(\phi(t_l^*)+\dot{\phi}(t_l^*)(t_l-t_l^*))}e^{(\gamma_l+i\dot{\phi}(t_l^*)(t-t_l)}\Theta(t_l-t). \label{eq:sinusoid2}
\end{align}
The expansion point $t_l^*$ is chosen to be near the cutoff time, $t_l^* =
t_l-\alpha T_l$, where $\alpha$ is a tunable parameter and the interval $T_l$
is the duration in which the approximation is valid:
\begin{align}
 |\frac{1}{2}\ddot{\phi}(t_l)T_l^2|=\epsilon < 1
\end{align}
and $\epsilon$ is a tunable parameter chosen to be to small. Equation
\eqref{eq:sinusoid2} implies that the coefficient $b_{0}$ for the $l$th
complex sinusoid is,
\begin{align}
  b_{0,l} &= A(t_l^*) e^{i(\phi(t_l^*)+\dot{\phi}(t_l^*)(t_l-t_l^*))}
\end{align}
and the frequency $\omega_l=\dot{\phi}(t_l^*)$.

In this paper, we chose the cutoff time $t_l$ of the first sinusoid to
correspond to the time at which the waveform has the highest frequency
detectable by the LIGO detector band. The next sinusoid is chosen by moving to
an earlier time, \mbox{$t_{l+1} = t_l - T_l$}. Since we want the $l$th
sinusoid to be mostly present on the interval \mbox{$t_{l}-T_l < t < t_l$}, we
choose the damping factor to be \mbox{$\gamma_l = \beta / T_l$}, where $\beta$
is a tunable parameter. This procedure is repeated until the time $t_l$
corresponds to a time in the waveform that has frequency below the LIGO
detector band. Hence the number of sinusoids is dependent on the value of
$\epsilon$, the rate of frequency change ($\ddot{\phi}(t)$), which is
dependent on the masses of the system, and the detector bandwidth. For more
information on this procedure, see \cite{Luan2011}.

\begin{figure}[htb]
  \centering
  \includegraphics[width=\columnwidth]{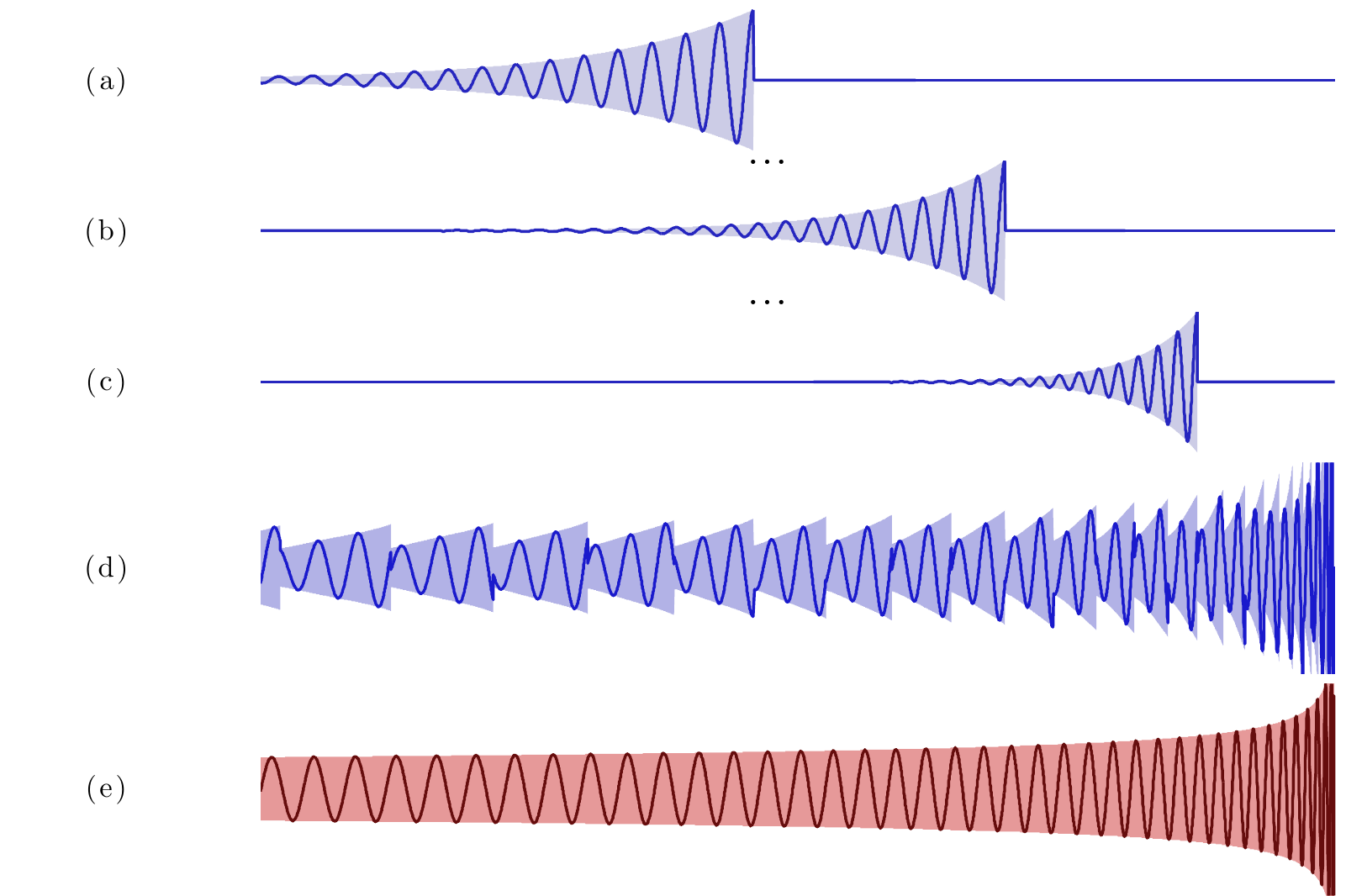}
  \caption{An illustrative diagram demonstrating the ability to linearly
      sum exponentially increasing constant frequency sinusoids to approximate
      an inspiral like waveform. The top three panels (a-c) show three example
      sinusoids with different damping, frequency and cutoff time
      factors. Panel (d) shows the linear addition of all the sinusoids (at
      different scales). Panel (e) shows the exact inspiral-like
      waveform. Note that this figure is only for illustrative purposes.}
  \label{fig:iirdemo2d}
\end{figure}

We can now approximate the phases \mbox{$\hat{h}(t) =
  A(t)e^{i\phi(t)}$} by an addition of a series of damped sinusoids $u(t)$
with cutoff times $t_l$:
\begin{align}\label{eq:approx}
  A(t)e^{i\phi(t)} \simeq U(t) &= \sum_l u_l(t) \nonumber \\
  &= \sum_l b_{0,l} e^{(\gamma_l + i
    \omega_l)(t-t_l)}\Theta(t_l-t).
\end{align}
Figure \ref{fig:iirdemo2d} shows an illustration of how damped constant
frequency sinusoids can add to give an inspiral like waveform.

\subsection{Summed Parallel IIR filtering}

Each complex sinusoid $u_{l}(t)$ in equation \eqref{eq:approx} can be searched
for in the data $x$ using the single pole IIR filter \eqref{eq:IIRdiff}. Here
the cutoff time is incorporated by running each filter on a delay, $d_l = t_l
/ \Dt$. The output of the $l$th filter at time $k$ is
\begin{align}
  y_{k,l} = a_{1,l} y_{k-1,l} + b_{0,l} x_{k-d_l}.
\end{align}
The linear summation of the output of \emph{all} filters is the
cross-correlation of the data $x$ and the approximate waveform $U(t)$ in
\eqref{eq:approx}:
\begin{align}
  z_k \simeq 2\Dt \sum_l y_{k,l}. \label{eq:zfilter}
\end{align}
Here $z$ is equivalent to the value computed by the discrete time domain two
phase filter \eqref{eq:zdiscrete} when using a template \mbox{$\hat{h}(t) =
  U(t)$}. From equation \eqref{eq:rho}, it follows that the absolute value of
the summation \eqref{eq:zfilter} divided by $\sigma_U$ is the SNR, which we
term the output of the \emph{Summed Parallel Infinite Impulse Response}
(SPIIR). The normalisation factor $\sigma_U$ is defined as
\begin{align}
  \sigma_U^2 = 4\halfint \frac{\left| \tilde{U}_{\Re}(f)\right|^2}{S_n(f)}df.
\end{align}
Where $\tilde{U}_{\Re}(f)$ is the Fourier transform of the real part of
$U(t)$, (which approximates $h_c(t)$). The similarity of the SPIIR output and
the matched filter output will depend on how well $U(t)$ approximates the
given template.

\section{Implementation for Performance Testing}\label{sec:method}

\subsection{IIR bank construction}

To confirm the ability of the SPIIR method to recover a good SNR, it is first
required to show that the approximate inspiral waveform \eqref{eq:approx} is a
good ``match'' to the theoretical inspiral waveform \eqref{eq:waveform}. We
define the \emph{overlap} $\Delta$ as the inner product of the approximate
waveform $U$ and the template $h$:
\begin{align}
  \Delta =
  \inner{\frac{h}{\sqrt{\inner{h}{h}}}}{\frac{U}{\sqrt{\inner{U}{U}}}} =
  \frac{\inner{h}{U}}{\sqrt{\inner{h}{h}\inner{U}{U}}}
\end{align}
We initially approximate a canonical 2PN 1.4-1.4 $M_\odot$ inspiral waveform
band limited to 10-1500\unit{Hz} using the value of the tunable parameters
$\epsilon$, $\alpha$ and $\beta$ to be consistent with the high overlap
results of \cite{Luan2011}. With some minor variation of their values, we aim
to recover the highest overlap possible. Once a good choice of $\alpha$ and
$\beta$ is found for the 2PN 1.4-1.4 $M_\odot$ template, we use the same
values for other templates, but vary the value $\epsilon$ (and consequently
the number of IIR filters in each bank) to see the effect on overlap.

\subsection{Detector Data Simulation} \label{sec:simulation}

To test the detection efficiency of the SPIIR method compared to the frequency
domain matched filter, we will filter two mock signals, one for which the
input data is just LIGO-like noise, and the other with the same noise plus an
inspiral waveform injection scaled to represent a source at a chosen effective
distance $\Deff$.

For this test, we need to construct a finite segment of detector data to
filter. Because of the IIR filters should in principle should be run for an
infinite length of the input data, we need to run the IIR bank for a finite
``warm-up'' period before the output is consistent with that of an IIR filter
that has been running for an infinite amount of time. In practise, we choose
to run each filter for 2 $e$-foldings of time before we accept the output as
being identical to one which has run for an infinite amount of
time. Additionally, since each IIR filter in the bank runs on a delay, the
summed output \emph{of all the IIR filters} will not be produced until after
the longest delay time ($d_{\rm max}$) has passed. The filter that has the
longest delay ($d_{\rm max}$) is also the one that has the longest decay rate
$\gamma_{\rm max}$. In total, the input data must at least $d_{\rm max} +
2\gamma_{\rm max}^{-1}$ in length before any output is produced. Hence the
length of the input data is,
\begin{equation}
  N_{\rm input} = d_{\rm max} + 2\gamma_{\rm max}^{-1} + N_{\rm analysis}
\end{equation}
where $N_{\rm analysis}$ is the length of analysis period, which we choose to
be 4 seconds. Hence the 4\unit{s} SPIIR output will tell us whether there is
an injection that ended somewhere within those 4 seconds. At a sample rate of
4096\unit{Hz}, the analysis period is \mbox{$N_{\rm analysis}=16834$} data
points long. In our simulation, we find \mbox{$d_{\rm max} = 4081683$} and
\mbox{$2\gamma_{\rm max}^{-1} = 149432$}, resulting in \mbox{$N_{\rm
    input}=4247499$}.

\subsubsection{Noise generation}

The LIGO-like noise data is produced by creating a normally distributed white
noise time series of length $N_{\rm input}$, then colouring it by the
theoretical advanced LIGO noise spectrum $S_n(f)$ \ref{sec:PSD}. We then
over-whiten this time series using equation \eqref{eq:whiten} to produce the
waveform-free noise input data $x$:
\begin{align}
  x_{\rm noise}(t) = n^{\rm ow}(t).
\end{align}

\subsubsection{Waveform injection} \label{sec:inj}

We create our waveform injections by first producing an inspiral waveform
band-limited between 10 and 1500\unit{Hz}. The injection is padded with zeros
so that it has the length $N_{\rm input}$. The end of the waveform is chosen
so that it finishes somewhere after $d_m + 2\gamma_m^{-1}$ data points. The
injection signal is then over whitened using equation \eqref{eq:whiten}. The
over-whitened injection can then be placed in the over-whitened noise signal,
\begin{align}
  x_{\rm noise + injection}(t) = x_{\rm noise}(t) + h^{\rm ow}(t).
\end{align}

\subsubsection{Matched filter comparison}

As a comparison, we will also perform a frequency domain correlation matched
filter. For this process, since the input data is already over-whitened, it
only needs to be cross-correlated with the waveform. Section
\ref{sec:matchedfilter} outlines how this is done. The cosine phase
$h_c(t)$ gets pre-padded with enough zeros to get to length $N_{\rm
  input}$. This ensures that $\tilde{h}_c(f)$ has the same spectral resolution
as $\tilde{s}(f)$. The matched filter \eqref{eq:zcomplex} produces a time
series of $N_{\rm input}$ length. However the first $N_{\rm input}-N_{\rm
  analysis}$ data points are erroneous wrap-around caused by the FFT. Only the
interval $[N_{\rm input}-N_{\rm analysis} + 1, N_{\rm analysis}]$ is used to
determine if a waveform is present.

\subsection{Detection Efficiency}

To test the detection efficiency of the SPIIR method compared to the
traditional matched filter method we will construct several receiver operating
characteristic (ROC) curves for 2PN 1.4-1.4 $M_\odot$ waveforms injected for
different effective distances $\Deff$. To create each ROC curve, we
first find the false alarm rate. The false alarm rate is found by realising an
$N_{\rm input}$ length LIGO-like noise time series, filtering this input data,
and analysing the output of the 4\unit{s} analysis period (the SNR). We will
count this realisation as a false positive if at any point within the 4
seconds the SNR goes over a given SNR threshold. Several thresholds will be
chosen, giving the false positive as a function of threshold. After $>10^6$
noise realisations, the false alarm rate is simply the ratio of total number
of false positives to number of noise realisations. Likewise, to see if the
IIR filter doesn't miss too many true positives, we inject a 2PN 1.4-1.4
$M_\odot$ waveform using the prescribed method in \ref{sec:inj} for a given
$\Deff$ into LIGO-like noise. After filtering, if at any point within
the analysis period the SNR is above a given threshold, this realisation is
counted as a true positive. Again, after $>10^6$ noise realisations, we
calculate the detection rate as a ratio of the total number of true positives
to number of realisations. The plot of false alarm rate versus detection rate
gives the ROC curve.

\section{Results}\label{sec:results}

\subsection{Inspiral Waveform Overlap}

Starting with the canonical 1.4-1.4 $M_{\odot}$ second order post-Newtonian
binary waveform band limited to be between 10 and 1500 \unit{Hz} we found,
using the parameters $\epsilon = 0.04$, $\alpha = 0.99$, $\beta = 0.25$ in the
procedure outlined in Section \ref{sec:approx}, that can recover an overlap of
99\% using 687 IIR filters.

We find that increasing the value of $\epsilon$ will in general increase the
overlap, as the frequency space is more finely sampled. However there seems to
be a limit, as the damping factor $\gamma$ causes the adjacent IIR filters to
run into each other.

\begin{figure}[htp]
  \centering
  \includegraphics[width=\columnwidth]{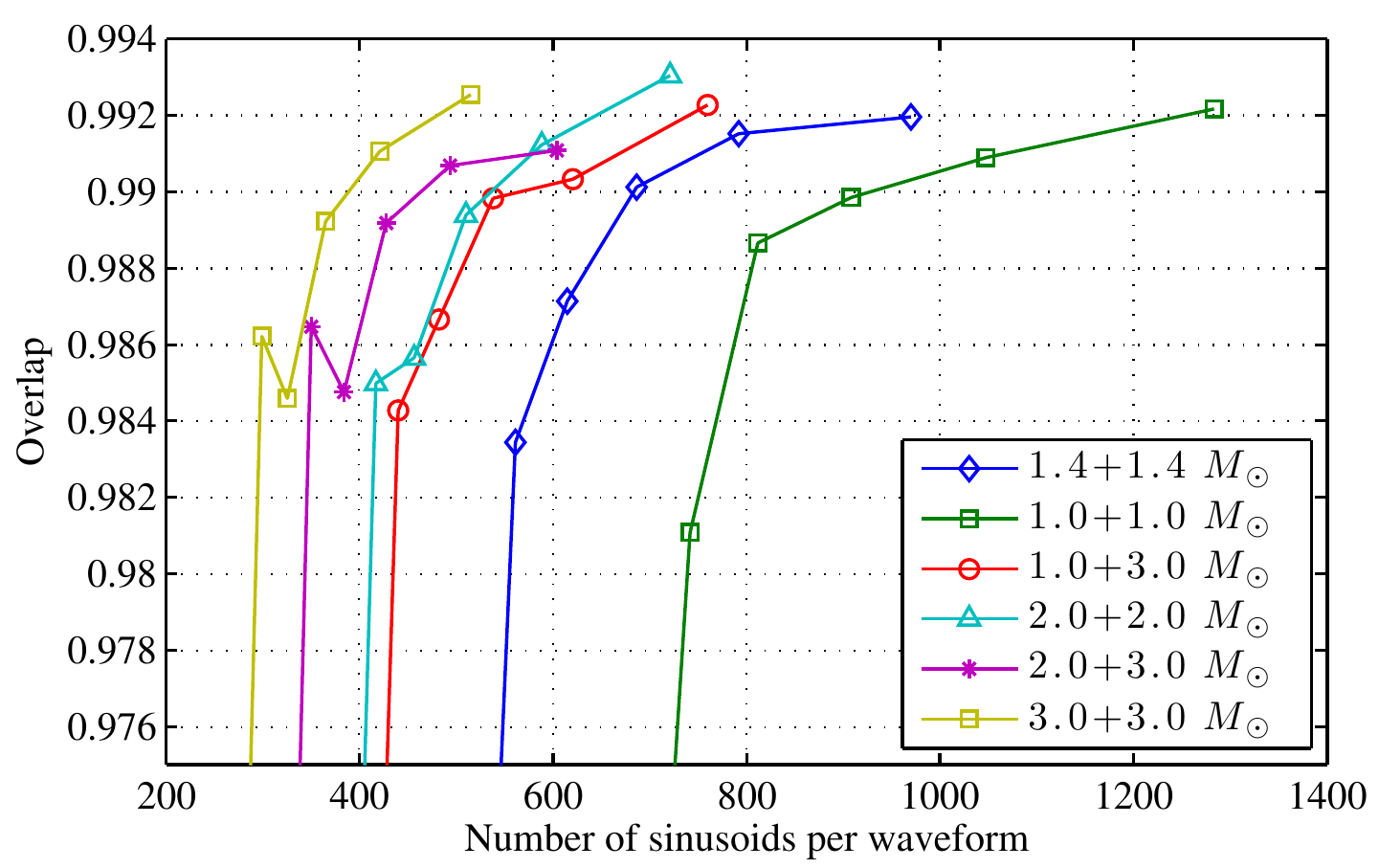}
  \caption{The overlap between the exact inspiral waveform and the approximate
    inspiral waveform as a function of number of damped sinusoids. In general
    the greater the number of sinusoids per waveform, the greater the
    overlap. However the choice of $\gamma_l$s greatly affects the overlap.}
  \label{fig:masspairs}
\end{figure}

With this choice of $\alpha$ and $\beta$ we are able to recover a high overlap
for different mass pairs as well. Figure \ref{fig:masspairs} shows the overlap
as a function of number of IIR filters for six different mass pairs.

\subsection{Ability to Recover SNR}

\begin{figure}
  \centering
  \includegraphics[width=\columnwidth]{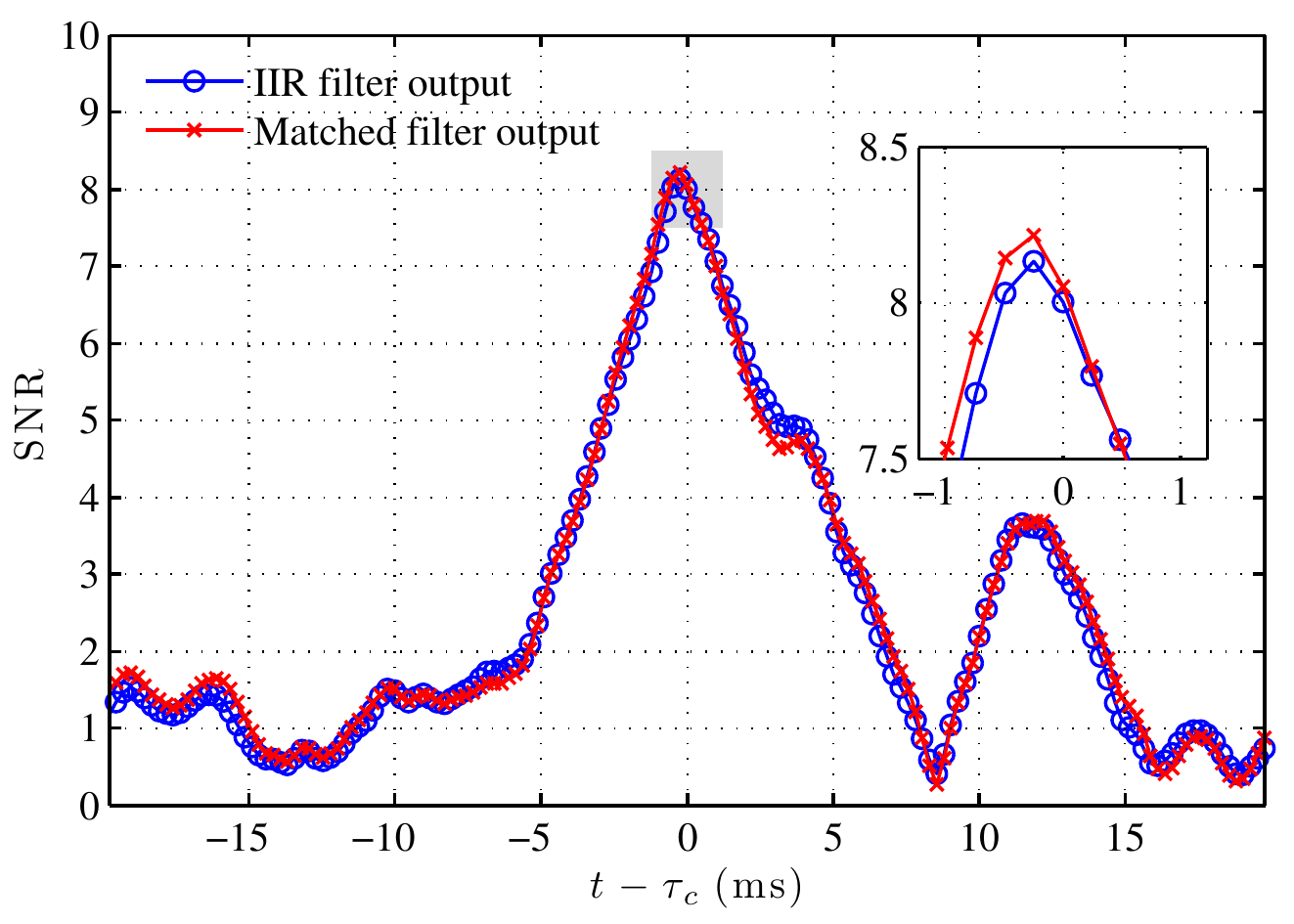}
  \caption{The SNR output of both the SPIIR method and a traditional matched
    filter method. The plot is centred on $t-\tau_c$ where $\tau_c$ is the
    time at which the injection ends. From the two curves, it is clear that
    the SPIIR method can return a very similar SNR to that from the optimal
    filter.}
    \label{fig:examplefilteroutput}
\end{figure}

Figure \ref{fig:examplefilteroutput} shows the SNR produced from both the
matched filter technique and the SPIIR method. The input time series is
constructed following Section \ref{sec:simulation}. The injection of a 2PN
1.4-1.4 $M_\odot$ waveform scaled for an effective distance of 250\unit{Mpc}
is added to LIGO-like noise. The $x$-axis of the plot is centred about end of
the injection ($t = \tau_c$), which is directly in the middle of the analysis
period. Around this time, the SNR peaks to 8.2, which is near the expected
value of 7.9 for an injection at this distance.
This plot shows that the SPIIR method is capable of recovering a very similar
SNR to the matched filter at all times.

\subsection{Detection Efficiency}

\begin{figure}[htb]
  \centering
  \includegraphics[width=\columnwidth]{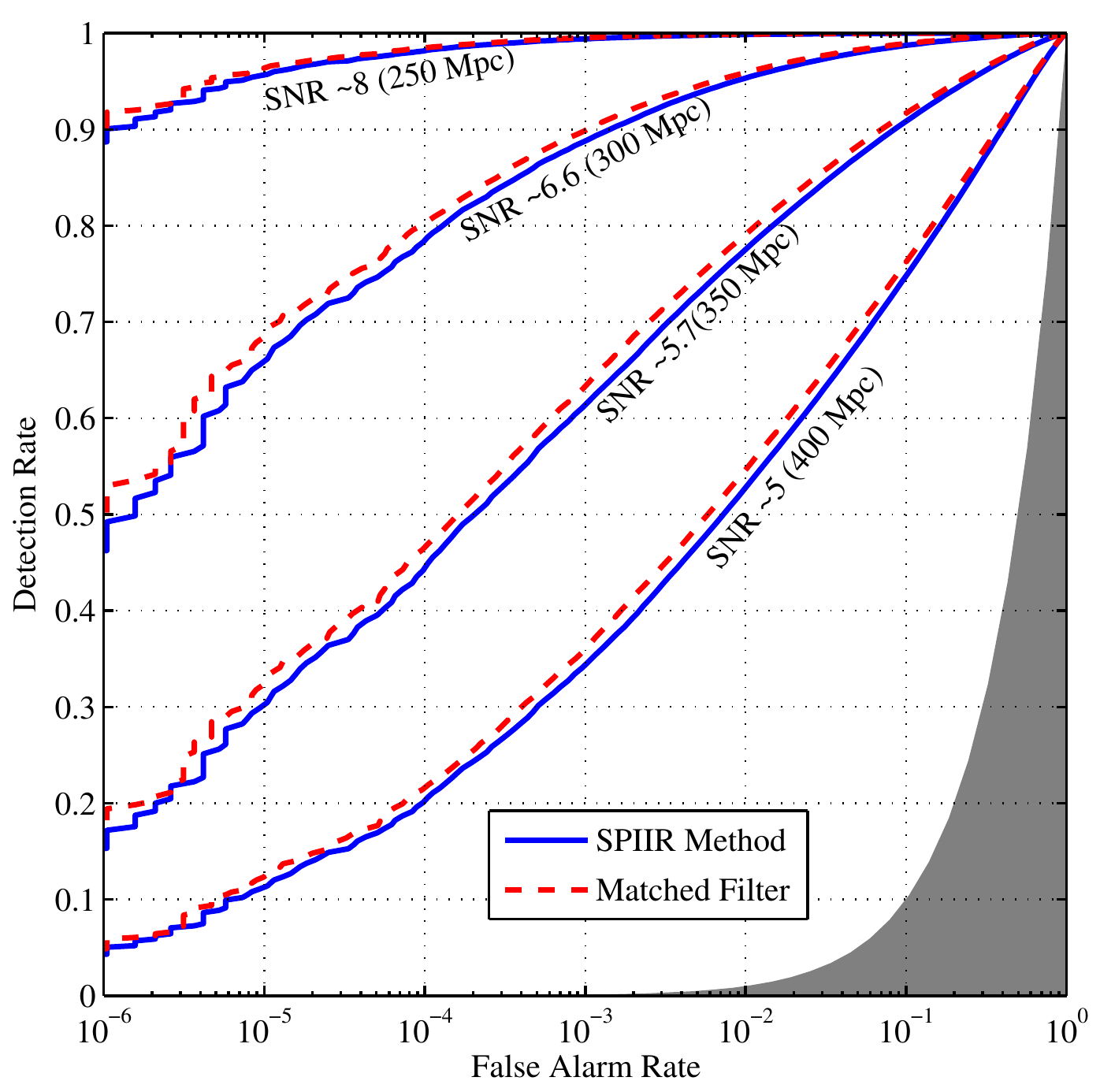}
  \caption{The receiver operating characteristics (ROC) of
     both the IIR filter method and the traditional matched filter method. The
     x-axis shows the false alarm rate, and the y-axis the detection rate. A
    one-to one relationship, which is the worst case scenario, is shown by the
    boundary of the shaded area. We show four different ROC curves, where the
    each curve represents the detection rate as a function of false alarm rate
    for waveform injected at effective distances of 250, 300, 350 and 400
    \unit{Mpc} (SNR $\sim$ 8, 6.6, 5.7 and 5 respectively).}
  \label{fig:roccurve}
\end{figure}

We analysed over $10^6$ independent noise realisations, for which the waveform
had been injected at $\Deff$ of 250, 300, 350, 400 \unit{Mpc}. We
performed both IIR filtering and traditional matched filtering. Figure
\ref{fig:roccurve} shows that the SPIIR method recovers most of the same
events as the traditional matched filter method. At false alarm rates of
greater than $10^{-5}$, the SPIIR method recovers greater than 99\% of the
injections recovered by the matched filter when searching for injections at an
effective distance of 250\unit{Mpc} (SNR$\sim$7.9). Even in the worst case, at
a false alarm rate of $10^{-6}$, the SPIIR method catches 4.5\% of injections
scaled at an extreme 400\unit{Mpc} (SNR$\sim$5), whereas the matched filter
catches 5\% of injections at this scale.

\section{Summary and Discussion}\label{sec:summary}

The use of a bank of simple IIR filters for each template as opposed to the
matched filter method enables us get two extra processes for a minimal
additional cost. The first is that the individual IIR filter outputs can be
arranged into groups, such that their total summed output is roughly
independent and orthogonal to each other. This enables, with minimal extra
overhead, the calculation of a $\chi^2$ distributed statistic, giving a
secondary method of verification. We will demonstrate this in an upcoming
paper. The second natural advantage of using a parallel bank of single-pole
IIR filters is that they can easily be executed in parallel using
multi-threaded processors, such as graphics processing units (GPUs). Indeed, a
side study has shown that this is possible \cite{Liu2011}. This leads to the
future possibility that a single personal computer may be able to process the
detection of GWs.

A further way to reduce the computation of the IIR calculation is to split the
incoming data into differently down-sampled channels. The output of each IIR
filter in the bank is the correlation of a fixed frequency sinusoid and the
incoming data. For the sinusoids that have frequencies $<$124\unit{Hz}, the
incoming data need only be sampled at 256\unit{Hz}. The current pipeline of
LLOID uses a similar multi-channel down-sampling in their detection
pipeline. Their pipeline consists of the integration of the open-source
real-time multimedia handling software \texttt{gstreamer} and the LIGO
Algorithm Library (LAL) \cite{gstlal}. This software library is an ideal
platform to integrate the SPIIR method. The total computation can also be
further reduced by sharing IIR filters (via interpolation) between different
templates \cite{Luan2011}.

Although the design of the IIR filter so far only applies to chirping,
post-Newtonian approximation inspirals, we have performed preliminary tests
using more complicated combinations of single-pole IIR filters to replicate
the waveform of an inspiral with spin. If the amplitude/frequency beating of a
spinning inspiral waveform can be simulated by the linear addition of two
different non-spinning inspirals with different masses, then it can be
approximated by a linear addition of damped sinusoids. In this case, the SPIIR
method can produce the SNR for the beating waveform. There is also the
possibility of using higher order IIR filters, although designing the
coefficients can be very difficult.

\section{Conclusion}

We have demonstrated that the through the use of a parallel bank of single
pole IIR filters, it is possible to approximate the SNR derived from the
matched filter with greater than 99\% overlap. The main advantage of our SPIIR
method is that it operates completely in the time domain, and in principle it
has zero latency (not taking into account whitening or computational
time). The SPIIR method recovers most of the injections the optimal matched
filter recovers.

We foresee that the use of IIR filters for time domain filtering of Advanced
LIGO will be ideal, as the waveforms will be much longer. The frequency domain
matched filter will take more time to calculate GW triggers, essentially
ruling out the possibility of triggering the detection the prompt optical
emission related to neutron star mergers (GRBs). We have shown that the use of
a parallel bank of IIR filters requires less computational cost, with minimal
detection rate loss, and most importantly can be calculated in the time-domain
with near zero latency.

\section{Acknowledgements}

We would like to thank Kipp Cannon, Drew Keppel and Chad Hanna for detailed
discussion on the design and implementation of low-latency detection
algorithms. This work was done in part during the LIGO Visiting Student
Researcher program, which was partially funded by the 2009 UWA Research
Collaboration Award. This research was supported by the Australian Research
Council. SH gratefully acknowledges the support of an Australian Postgraduate
Award.

\bibliographystyle{apsrev}
\bibliography{mycenbib}

\appendix
\section{Noise Spectral Density}\label{sec:PSD}
We use an algebraic expression for the noise spectral density of Advanced LIGO
detectors defined by,
\begin{align}
  \begin{split}
    S_h(f) = S_0\left\{  \left(\frac{f}{f_0}\right)^{-4.14} - 5\left(\frac{f_0}{f}\right)^2 + \right. \\
    \left. 111  \left(\frac{1 - \frac{f}{f_0}^2 + 0.5  \frac{f}{f_0}^4}{1. + 0.5\frac{f}{f_0}^2} \right)\right\};
  \end{split}
\end{align}
where, $f_0=215$Hz and $S_0 = 10^{49}$.

\end{document}